# Fractional Brownian Motions and their multifractal analysis applied to Parana river flow

M. N. Piacquadio Losada, R. Seoane, A. de la Barra, L. F. Caram

Facultad de Ingeniería, Universidad de Buenos Aires, Argentina)

**Abstract:**

A number of different analysis techniques have been used to analyze long-term time series data from different rivers, starting with the determination of the Hurst coefficient. We summarize the concept of fractals, multifractals and Fractional Brownian Motion (FBM), and apply some such techniques to daily stream flow data from the Paraná River recorded at Corrientes, Argentina, for 106 years. After determining the Hurst coefficient for the entire data set (H = 0.76), we analyze the data for each of four seasons and draw the corresponding FBM graphs and their multifractal spectra (MFS). Three of the seasons are similar, but autumn is very different for both FBM and MFS. Based on the MFS results, we propose a number of indices for measuring variations in stream flow, and determine the values of the indices for the three "similar" seasons. The indices are based on important parameters of the multifractal spectra $(\alpha, f(\alpha))$: $\alpha_{min}$, $\alpha_{max}$, $f_{min}$, $f_{max}$, and $\alpha(f_{max})$. The geometry of the spectra as well as the indices all indicate that Winter is the most stable season. This is in contrast to the Boxplot of seasonal stream flow data where Winter shows the largest variation. Thus, these indices provide insight into river flow stability, not detected in—and indeed contradictory to—that from basic statistical analysis.

## Introduction

There exists an interest in studying the properties of times series that describe hydrological processes such as precipitation and stream flows. This interest is justified not only to improve the mathematical description of these two parameters important for the hydrological cycle, but also for their application in the estimation of design parameters for hydrological projects, as well as their operation in a given river basin.

The response of stream flow to precipitation depends on the attributes of the river basin, which include its geographic location, the underlying geology, geomorphology, vegetation cover, as well as the precipitation.

A long series of empirical observations led Harold Edwin Hurst to formulate a statistical procedure for measuring the long-term persistence of time series applied to river flows (Hurst, 1951, 1956). Hurst proposed a statistic which would permit establishing design parameters for a dam to be built on the Nile River at Aswan. This statistic is determined by a range R (difference between maximum and minimum stream flows during a time interval τ) and a standard deviation S over the same period. The ratio R/S for different time intervals was defined by an empirical relationship:

$$\frac{R}{S} = \left(\frac{\tau}{2}\right)^H \qquad \text{Eq.1}$$

In honor of Hurst, the statistic is called the Hurst coefficient (or Hurst exponent), and is designated by the letter H as in Eq. (1).

In a paper dedicated to Hurst, Mandelbrot and Wallis (1968) compared Hurst's approach to traditional statistics, and proposed the notion of Fractional Gaussian Noise (FGN), within the context of Hydrology. In another paper, Mandelbrot and Van Ness (1968) proposed the concept of Fractional Brownian Motion (FBM) elaborated on the notion of fractional noises and what they denominated "Hurst's Law" for long-range dependence characterized by a value of H between 0.5 and 1.

There are a number of methods for estimating the Hurst coefficient, but this is not the subject of the present paper. See, e.g. Sánchez Granero et al. (2008); Trinidad Segovia et al. (2012); Torre et al. (2007).

A number of studies undertaking multifractal analysis of daily river flows reported in the literature (Pandey et al., 1998; De Bartolo et al., 2000) all use the thermodynamic algorithm, which forces the shape of the multifractal spectrum, as we show in this paper. Kantelhardt et al. (2002) proposed a different technique (multifractal detrended fluctuation analysis, MF-DFA), which has been applied by the same authors (e.g. Koscielny-Bunde et al., 2006) and by others including Zhang et al. (2008) for the Yangtze River. However, this technique does not lead to the generation of multifractal spectra. Rego et al. (2013) studies water levels not river flows. Moreover, they do not determine multifractal spectra, nor undertake seasonal analysis.

The objective of the study is to determine the Hurst coefficient of the stream flow of the Parana River, to conduct a Fractional Brownian Motion (FBM) and multifractal analysis for each season, and to propose a number of indices based on this analysis. We then compare the results and conclusions with those of traditional statistical data analysis.

Below, we present the "Study area and data" followed by a "Methods" section subdivided into Fractals, Multifractality, Thermodynamic formalism, and Fractional Brownian Motion (FBM). Results are also presented in a number of subsections, followed by the Conclusions.

## Study area and data

The Parana River basin has a continental scale: it is the second largest river in South America, originating in Brazil and draining Paraguay and a part of Argentina, before reaching the Atlantic Ocean as River Plate (Figure 1). The study is based on river flow data near Corrientes city, in Argentina. Figure 2 shows the data from 1904-2010. The Paraná River basin up to Corrientes has an area of 1.85 million km$^2$, and its tributaries include the following rivers: Iguazú, Bermejo. Pilcomayo, and Paraguay.

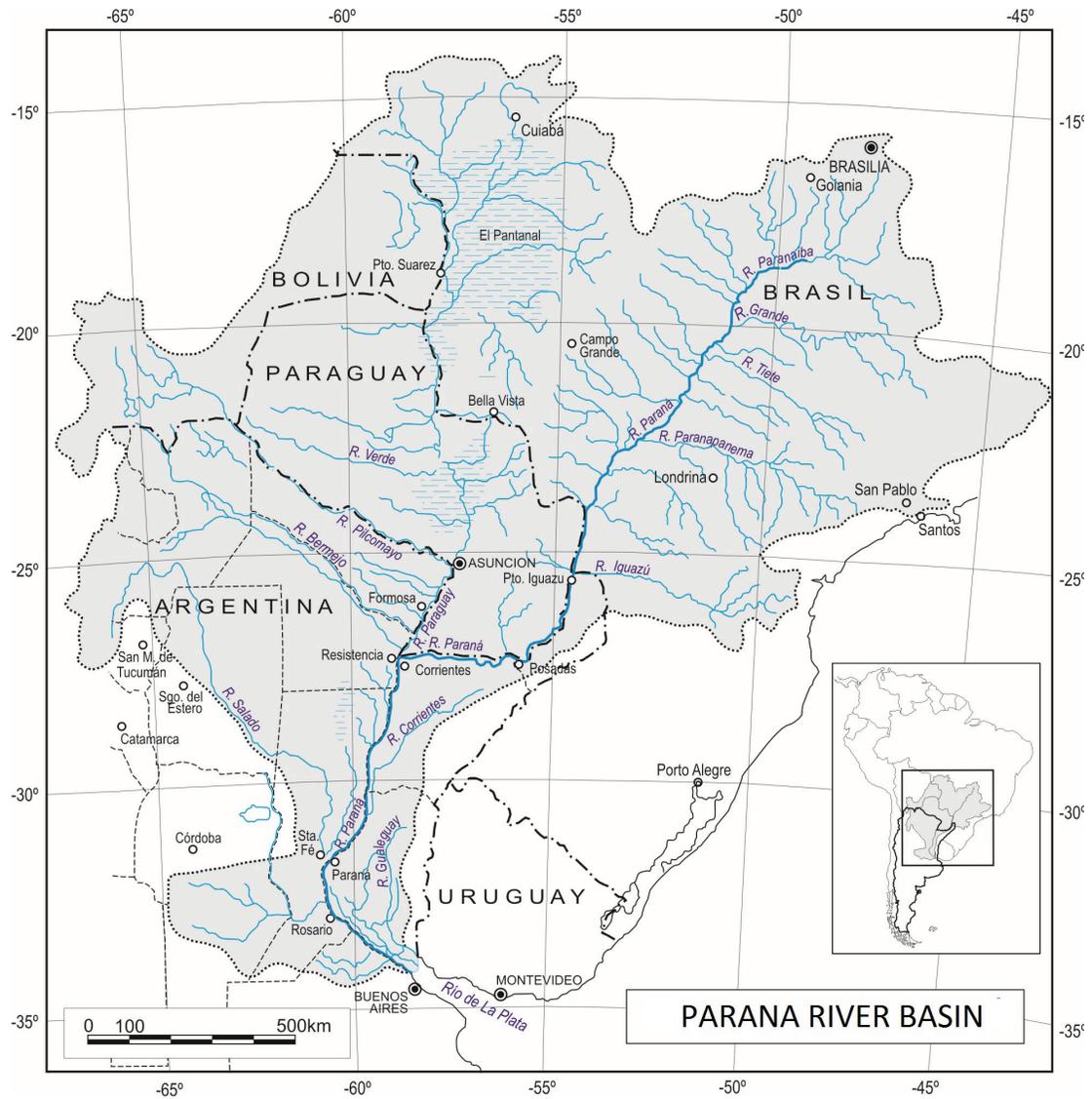

**Figure 1. Parana river basin**

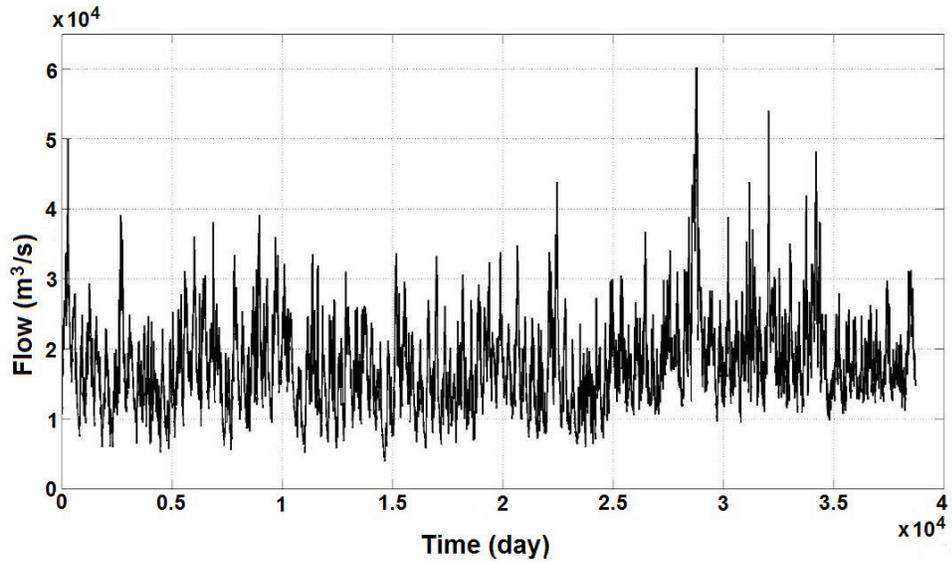

**Figure 2. Daily stream flow for the Paraná River, Corrientes, Argentina, 1904-2010.**

The Parana River flow at Corrientes shows strong seasonal variations as shown in Figure 3, with Boxplot for each season shown in Figure 4.

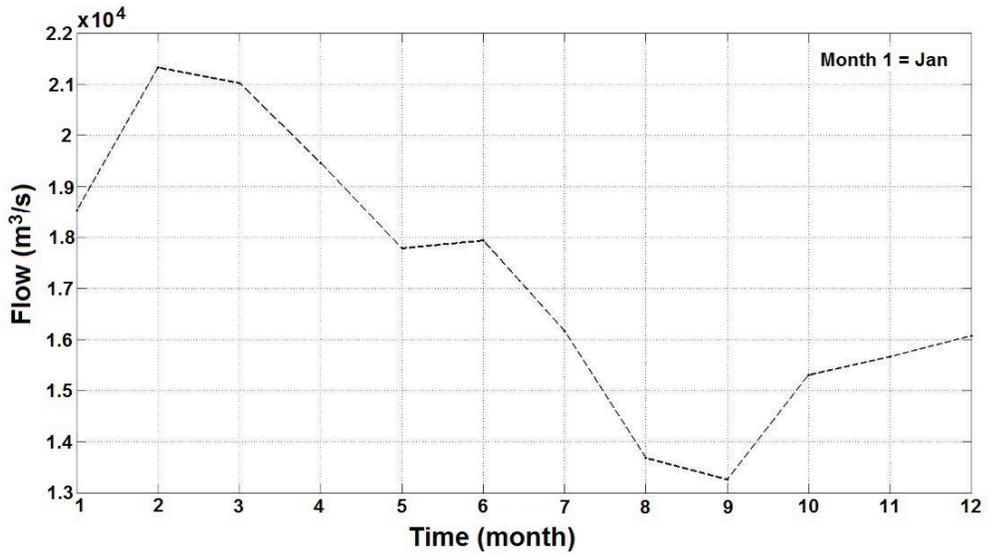

**Figure 3. Monthly average stream flows for the Paraná river, Corrientes, 1904-2010.**

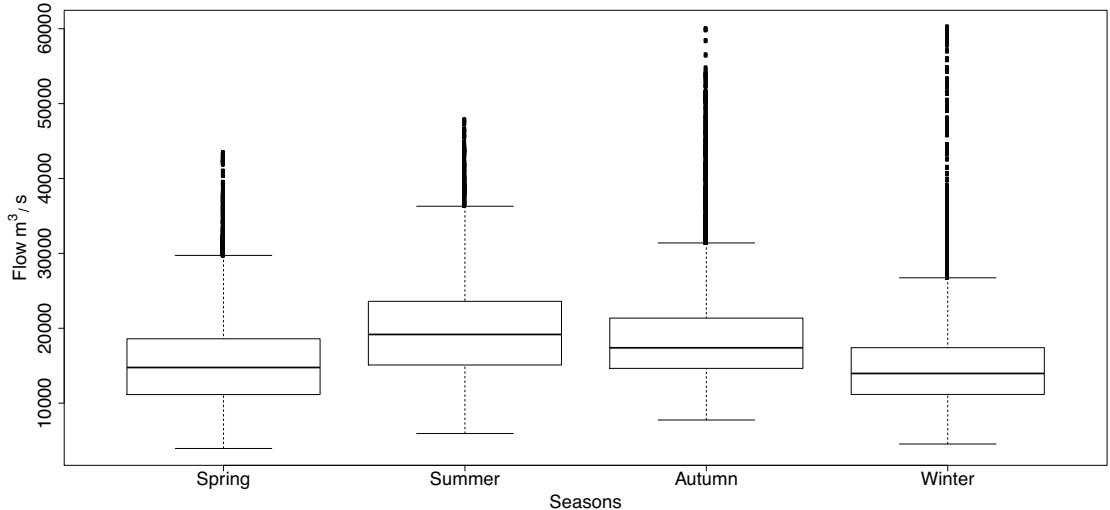

**Figure 4. Boxplot showing means and quartiles of stream flow (m³/s) at Corrientes, for the four seasons, 1904-2010.**

Stream flows are lowest in Winter, but with large range of extreme values. Autumn also shows very large range with a higher average value. The inter-quartile range (inner boxes) shows relatively small variation across seasons: it is smallest for Winter, somewhat larger for Spring and Autumn, and largest for Summer. The annual cycle of Paraná River flow at Corrientes coincides with data analyzed by Camillioni and Barros (2000): for the period 1930-80 flow was higher in Summer (January) decreasing to September.

## Methods

Below we define some basic mathematical concepts on fractals, multifractals and Fractional Brownian Motion.

**Fractals**

A fractal is an object of not-necessarily integral dimension $d$ in the sense of Hausdorff ($d_H$, see, e.g. Falconer, 2003, Sec. 2.2) or Minkovsky ($d_B$ or Box dimension, see Falconer, 2003, Sec. 3.1). We work in $\Re^2$.

We will use $d_B$ throughout this paper. We cover the object under study with a grid of square boxes of side $\ell$. We count the number $N(\ell)$ of the boxes covering the object under study. In Fig. 5 the object is a square of side 1.

Notice that there is a relationship between $N(\ell)$ and $1/\ell$: $N(\ell) = (1/\ell)^2$. This "square" exponent is the dimension of the object. We need, in general, a covering mesh of boxes of the same size and shape.

Another example is the von Koch snowflake (Fig. 6) whose geometry is best understood using a triangular net of equilateral triangles of size $\ell = 1/3, 1/3^2,\ldots$ .

The condition "$\ell \rightarrow 0$", so often hand-in-hand with any definition of dimension, $d_H$ or $d_B$, should be replaced hereafter with "$\ell$ as small as it makes sense in order to calculate $d_B$" (see below for some key examples).

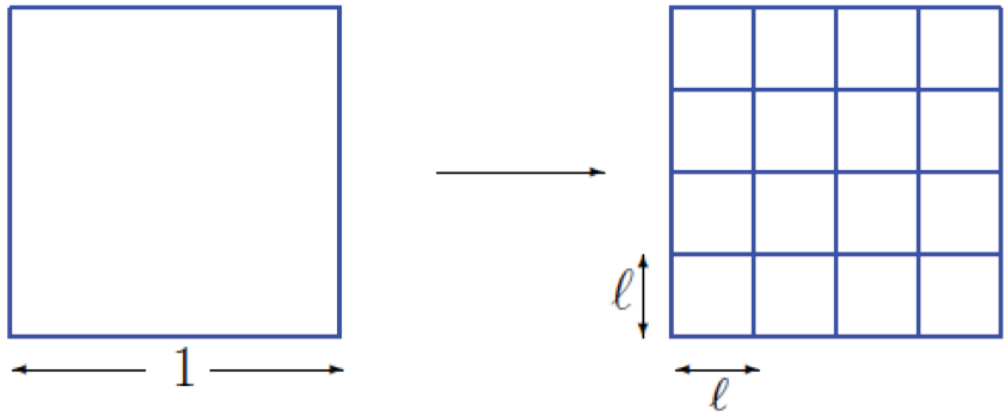

**Figure 5. Covering a square with square boxes of size $\ell$**

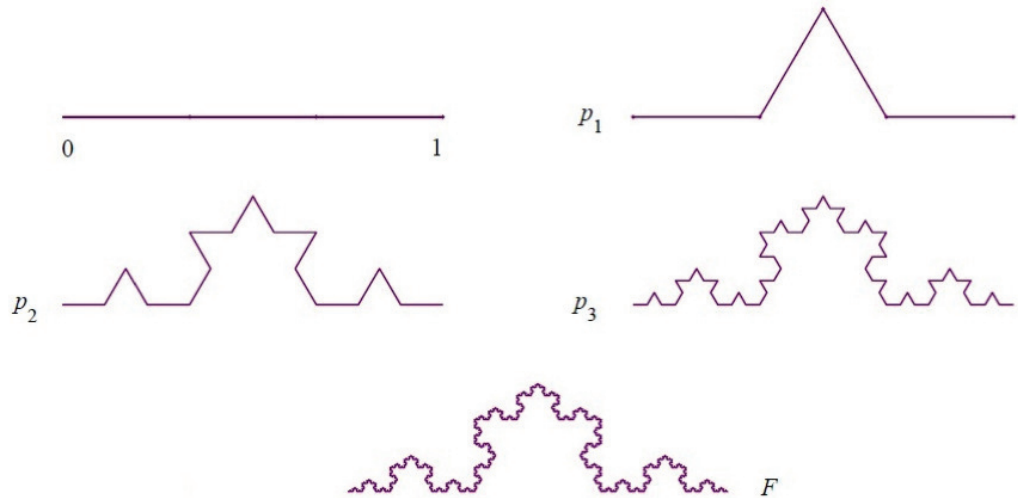

**Figure 6a. Steps in the generation of the von Koch snowflake**

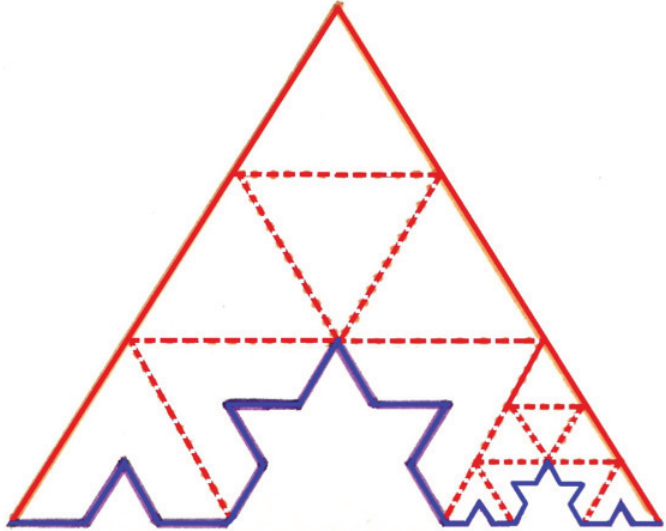

**Figure 6b. Von Koch snowflake covered with triangles**

Notice that the notion of self-similarity applies here. For instance, in Fig. 6b, the subfractal with horizontal base [2/3, 1] is identical, via a change of scale with the whole fractal *F*.

**Multifractality**

We start with a simple example: the Cantor ternary, with unit weight or measure, uniformly distributed (Fig. 7).

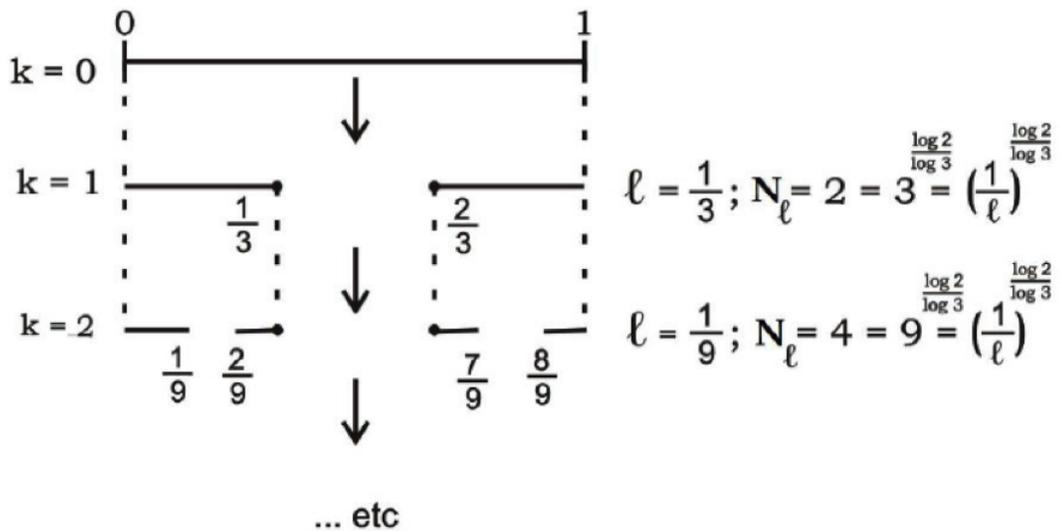

**Figure 7. Steps in the generation of the Cantor set.**

In the $k^{th}$ step of its construction we have $2^k$ "boxes" ($B$) of size $\ell = 1/3^k$ and weight or measure $1/2^k$.

The density $\delta = \frac{\text{weight}}{\text{length}} = (\frac{1}{2^k})/(\frac{1}{3^k}) = 3^k/2^k = (3/2)^k \to \infty$ if $k \to \infty$, or equivalently, $\ell \to 0$. Hence, it makes no sense to consider the density $\delta$.

The log-log version of $\delta$ is called "concentration":= $\alpha(B_i) = \frac{\ln \text{weight}}{\ln \text{length}} = \ln 2/\ln 3$ and is well defined for all boxes $B_i$.

But if the weight or measure is not uniformly distributed, then $\alpha(B_i)$ may vary from box to box. In that case, let $\Omega_\alpha$ be the set of all boxes with the same α-concentration, or rather, all elements with the same α. Then, by definition, $f(\alpha) = d_B(\Omega_\alpha)$, that is, if $N(\alpha)$ is the number of elements in $\Omega_\alpha$, then $f(\alpha) = \frac{\ln N(\alpha)}{\ln(1/\ell)}$.

The curve $(\alpha, f(\alpha))$ is called the multifractal spectrum of the unit weight or measure $p$ over the Cantor set. In the same sense, we can obtain the spectrum $(\alpha, f(\alpha))$ of any unit weight or measure $p$ defined on a certain fractal $F$.

It is expected that such spectra are usually curves with $f''(\alpha) < 0$, i.e. a curve in the form of an inverted U. But examples from data taken from a real, physical problem, can produce an $(\alpha, f(\alpha))$ spectrum of several disconnected curves, different from one another, but each one with negative second derivative: see, e.g. Francois, Piacquadio and Daraio (2011); Salvo and Piacquadio (2016). …And yet, many authors consider all multifractal spectra to be given by the so-called Thermodynamic Algorithm (where the spectrum is a single curve with $f''(\alpha) < 0$, always) to be a correct substitution for the multifractal spectrum defined above.

Let us briefly describe the thermodynamic formalism as a multifractal spectrum.

**The thermodynamic formalism as a multifractal spectrum of a fractal**

Let $F$ be a planar fractal, endowed with a probability measure $p$, and superposed with a grid of square boxes $B_i$ of side $\ell$.

Let $p_i$ be the measure of the fractal inside box $B_i$.

Let $q \in \mathfrak{R}$, $q \in (-\infty, \infty)$ a parameter.

By definition $\tau(q) = \frac{\ln \sum_i [p(B_i)]^q}{\ln \ell}$; we want to obtain $\alpha$ and $f(\alpha)$ of $F$ via $\tau(q)$.

We start working with the $\Sigma_i$ involved in $\tau(q)$.

We know: $\alpha_i = \frac{\ln p_i}{\ln \ell}$, then $\ln p_i = \alpha_i \ln \ell = \ln \ell^{\alpha_i}$, hence $p_i = \ell^{\alpha_i}$.
Therefore $\Sigma_i = \Sigma_i \ell^{\alpha_i q} = \Sigma_\alpha \ell^{\alpha q} N_\alpha$, where $N_\alpha$ is the number of boxes with concentration roughly equal to $\alpha$. We know that $f(\alpha) = \frac{\ln N(\alpha)}{\ln(1/\ell)} = -\frac{\ln N(\alpha)}{\ln(\ell)}$, $\ln \ell^{f(\alpha)} = -\ln N_\alpha$, hence $\ell^{-f(\alpha)} = N_\alpha$.

Then $\Sigma_i = \Sigma_\alpha \ell^{\alpha q - f(\alpha)}$. It is thought that $\Sigma_i = \Sigma_\alpha \ell^{\alpha q - f(\alpha)} \sim \ell^{\min_\alpha (\alpha q - f(\alpha))}$ if $\ell$ is very small. Therefore, $\tau(q)$ can be estimated by:

$$(\ln \ell^{\min \alpha(\alpha q - f(\alpha))})/\ln \ell = \min_\alpha (\alpha q - f(\alpha))$$

This implies $d(\alpha q - f(\alpha))/d\alpha = (q - f'(\alpha)) = 0$, and $\frac{d^2(\alpha q - f(\alpha))}{d\alpha^2} = \frac{d(q - f'(\alpha))}{d\alpha} > 0$.

From the first condition we obtain $q = f'(\alpha)$; from the second one we have $-f''(\alpha) > 0$, so $f''(\alpha) < 0$. Notice that, for the thermodynamic formalism, $f''(\alpha)$ **will always be negative**, so that $f(\alpha)$ has an inverted U shape. Such a shape does not allow for bimultifractal spectra, as in Salvo and Piacquadio (2016), François, Piacquadio, and Daraio (2011), and further below in this paper.

It is for this reason that we work, consistently throughout this paper, with the ***definition*** of $(\alpha, f_B(\alpha))$, by the box-counting method.

**Fractional Brownian Motion (FBM)**

We start with an observation: the fractality of the von Koch snowflake is implied by progressive "creasing": first there is a triangular crease in the center of the unit segment (see Fig. 6a); then we have four new smaller creases, and we notice that two of these are at left and right, respectively, of the initial triangle. This up-and-down-and-left-and-right creasing is the very cause of the fractal character of the snowflake.

Let us now consider a ***function***—which the snowflake is definitely not—$f(t)$, which can only evolve from $f(t)$ to $f(t + \delta)$ going only ***up*** or ***down***. In order to think of such a function as a fractal, it is necessary for it to be "infinitely creased". For instance, from $f(t)$ to $f(t + \varepsilon)$ to $f(t + 2\varepsilon)$ we should have a zig-zag: "zig" from $f(t)$ to $f(t + \varepsilon)$, "zag" from $f(t + \varepsilon)$ to $f(t + 2\varepsilon)$. Any "zoom in", for instance, in the "zig" from $f(t)$ to $f(t + \varepsilon)$ will reveal smaller and smaller zig-zags (as we zoom in and in, *ad infinitum*). Needless to say, in a practical case, such "ad infinitum" cannot take place. Such a function is called a Fractional Brownian Motion (FBM).

Observation:

a) Going back for a moment to planar fractals, the infinite creasing of a curve (such as the snowflake) is a necessary condition for it to be a fractal, but not a sufficient one for it to have a ***definite $d_B$*** ($\ln 4 / \ln 3$ in the case of the snowflake).
   The other ***key*** condition is the ***self similarity*** already spoken of.
b) For a function $f(t)$ to be a fractal, it is necessary to be infinitely creased… but in order to have a ***definite $d_B$***, the other ***key*** condition is ***self affinity***: that is, that a number $H \in (0,1)$ exists, such that $|f(t + \Delta) - f(t)| \sim \Delta H$ on average, $\Delta$ a real number. Such an H is called "the Hurst coefficient or exponent" of the FBM, and it is thought that $d_B(F) = 2 - H$, where $F$ is the graph of the function $f(t)$.

One way of determining the Hurst exponent of the FBM function is the so-called rescaled range method $(R/S)$ as proposed by Hurst (1951) and Mandelbrot and Wallis (1968). The range $R$ is the difference between the maximum and the minimum value of the function for a given period; $S$ is its standard deviation over the same period.

**Results**

**The FBM associated with the flow of the Parana River**

We first model the daily stream flow of the Paraná River over 106 years (1904-2010) (Figure 2) as an FBM function. We start with estimating the Hurst coefficient, following the traditional R/S method, i.e. from the slope of the best-fit line log (R/S) vs. log (n), where n is the number of data points. The corresponding value of $H$ is approximately 0.76, as shown by the slope in Figure 8. The geometric determination of H allows a visualization of its accuracy not found in purely computational methods.

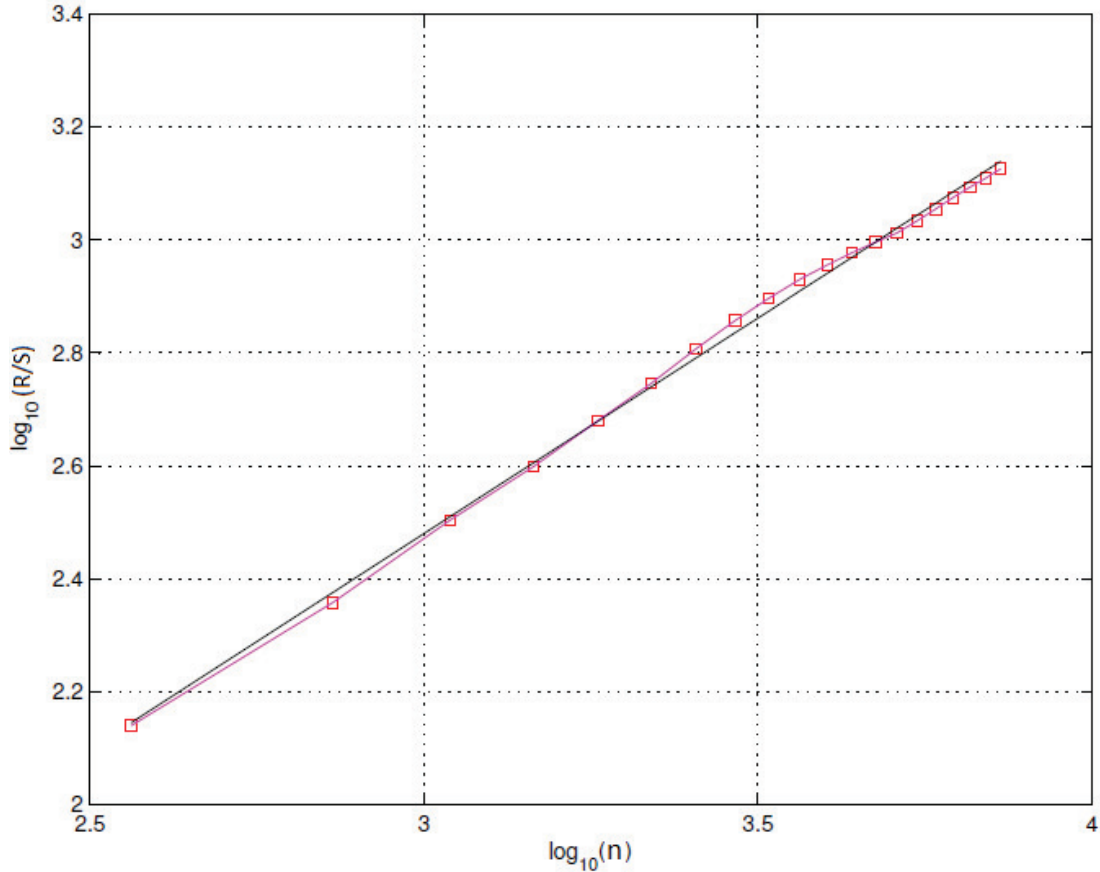

**Figure 8. Hurst coefficient estimate of the Parana river, Corrientes, Argentina, 1904-2010, using the R/S method**

Hurst (1951) found H values in the neighborhood of 0.75 for a wide range of phenomena. Rago et al. (2013) determined H to have values from 0.65 to 0.78, for a number of Brazilian rivers. For 41 rivers in different parts of the world, Koscielny-Bunde et al. (2006) found values ranging from 0.55 to 0.95. They note that "*there is no universal scaling behavior since the long-term exponents vary strongly from river to river and reflect the fact that there exist different mechanisms for floods where each may induce different scaling*".

**Seasonal analysis**

We next propose to study the stream flow corresponding to different seasons by extracting the 106 Winter solstices, the 106 Summer solstices and the Autumn and Spring equinoxes from the total T of days in the 106 years.

Our assumption, for the moment, is that the Hurst coefficient estimated from daily data over 106 years (365 x 106 = approximately 38000 data points), $H \sim 0.76$, is inherited by the 106 solstices and equinoxes. Below we refer to the solstices and equinoxes simply by the name of the corresponding season.

The 106 Winter flows are the $X_{1,\ldots,106}$ of the R/S method. Note that $Y_i = X_i - \overline{X_{1,\ldots,106}}$ ($i$ goes from 1 to 106) whereas $Z_i = \sum_{k=1}^{i} Y_k$.

We rescaled with the usual $(1/N, 1/N^H)$ factors, where $N = 106$ and $H \sim 0.76$; that is, the horizontal (time) variable will go from $\frac{1}{N} = \frac{1}{106} \sim 0$ to unity, whereas each vertical $Z_i$ will be divided by $106^{0.76}$. The rescaled values are called $Z(t)$ or $Z_i$. This will yield a Winter graph.

We repeat the same procedure with the other seasons, obtaining four graphs, see Figure 9a and Figure 9b.

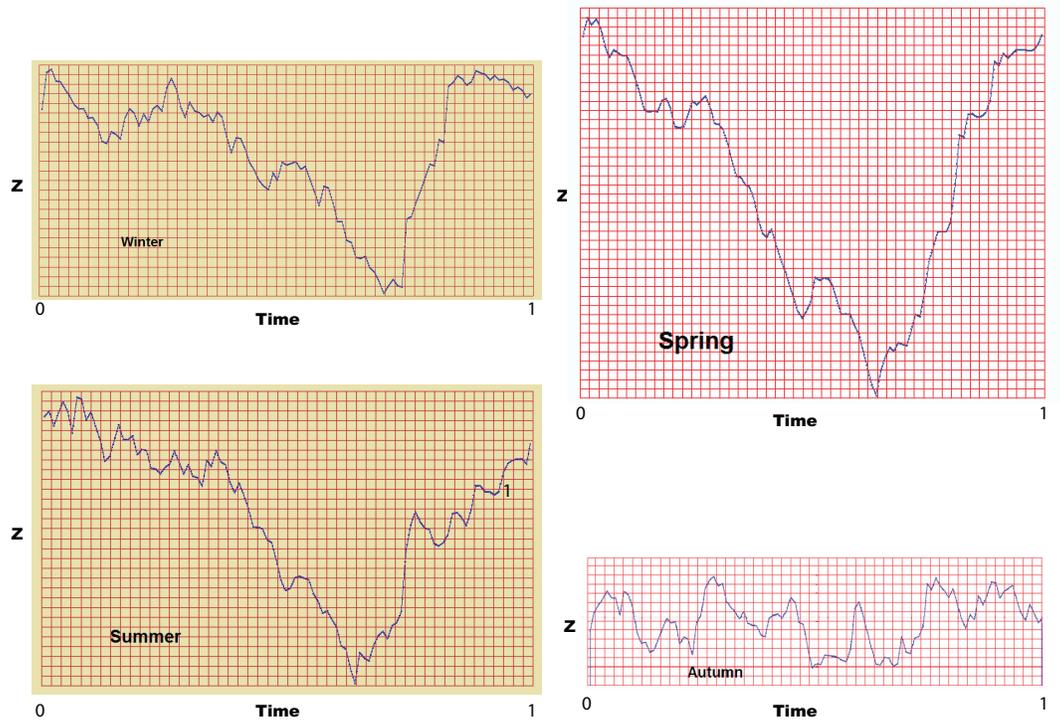

**Figure 9a. The FBM graphs, represented by the corresponding rescaled Z values against time, for each season**

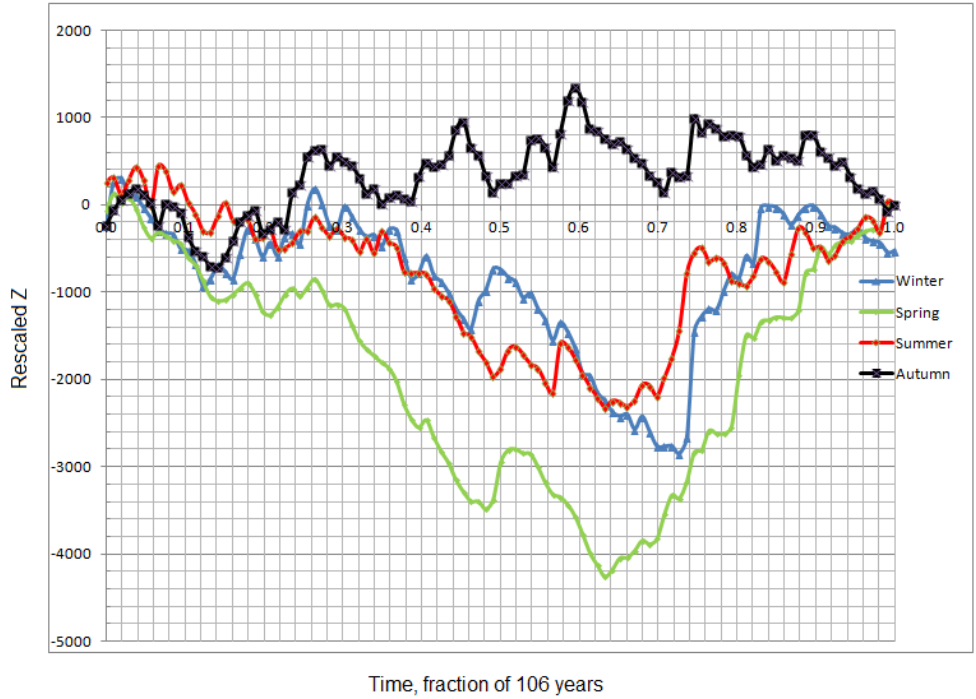

**Figure 9b. The FBM graphs, represented by the corresponding rescaled Z values against time, for the four seasons together.**

For three of the four seasons, the rescaled Z values decline up to a certain point (1978 for Winter and 1972 for Spring and Summer) and increase thereafter. There is no such pattern for the Autumn data. No long-term trend is visible in the daily stream flow data shown in Fig. 2. Two major dams on the Parana River in Brazil upstream of Corrientes (the measurement point) were completed in the 1970s: Jupia in 1974 and Ilha Solteira in 1978. A provisional dam was made at the very large Itaipú power project in 1978.

The vertical range of rescaled Z values is smallest for Autumn, without the long-term trends seen in the other three seasons. For these three, the range is smallest for Summer, followed by Winter and finally by Spring. These observations do not emerge in the boxplot of seasonal flow (Fig. 4), where the inter-quartile ranges (the inner boxes) for all seasons are about the same. Indeed, the extreme values ("whiskers") in the boxplot are seen for Winter, suggesting greater variability in this season.

Next, we propose to obtain the multifractal spectrum $(\alpha, f(\alpha))$ of each of the four seasons, to represent these spectra together, and to draw conclusions from their differences.

Let us consider one of these rescaled Z graphs (Fig. 9). We cover it with a grid of square boxes of size $\ell$, and proceed exactly as in Sec. "Multifractality", with one modification: $\Omega_\alpha$ will be the number of boxes with concentrations *roughly* equal to $\alpha$, i.e. $[\alpha - \Delta\alpha/2, \alpha + \Delta\alpha/2]$, for an appropriate $\Delta\alpha$. This "vagueness" takes care of the case in which all $\alpha(B_i)$ are different for different values of $i$. In that case, the answer to "how many elements are in $\Omega_\alpha = \{\text{all concentrations} = \alpha\}$?" will be 1, so $N_\alpha = 1$, and $d_B(\Omega_\alpha) = \frac{\ln N(\alpha)}{\ln(\frac{1}{\ell})} = 0$!

Notice that the $d_B$ dimension of each FBM graph will be $\frac{\ln(\#of\ \textbf{\textit{all}}\ \text{boxes that cover the graph})}{\ln(1/\ell)}$.

The choice of $\ell$ and $\Delta\alpha$ for a reasonably shaped multifractal spectrum $(\alpha, f(\alpha))$ of a fractal (an FBM in this case) is entirely artisan like (see, e.g. Serinaldi, 2010, Section 5): by trial and error we found that, for $\ell = 1/50$ and $\Delta\alpha = 0.08$ **all** four multifractal spectra of the different seasons are reasonably shaped, and can be compared.

We agree with the usual result: $2 - d_B(graph) = H_{graph}$

## Characteristics of $d_B$ of an FBM spectrum

Recall that the total weight or measure of each graph (or of any fractal under study) is unity. Therefore, all measures of all boxes are in the unit interval.

### $\alpha_{min}$

If we take a very heavy box, with normalized weight or measure $p$ near 1, but $p < 1$, we obtain $\ln p$ negative and very small. Now $\alpha := \ln p / \ln \ell$, but $\ln \ell$, negative, is the same for all boxes, a negative constant, which implies that $\alpha = \ln p / \ln \ell$ is positive and smaller than other boxes.

Therefore, the heaviest boxes correspond to $\alpha_{min}$.

For all four spectra, $\alpha_{min} = 1$, for our choice of $\ell$ and $\Delta\alpha$.

Since $\alpha_{min} = 1$ and $\Delta\alpha = 0.08$, let us focus on the Winter graph in order to explain the method that yields the spectrum.

The horizontal α-axis is divided into $\Delta\alpha$ intervals: (1, 1.08]; (1.08, 1.16]; (1.16, 1.24]… etc. In the interval (1, 1.08] we have 16 boxes, so the corresponding $f(\alpha)$ is $\ln 16 / \ln 50$; in the interval (1.08, 1.16] we find 20 boxes, with an $f(\alpha) = \ln 20 / \ln 50$,… etc.

### $f_{max}$

Another characteristic of a multifractal spectrum: adding all boxes in all the intervals we have, see above, $\ln(\text{no. of boxes}) / \ln 50 = d_B$ (of the fractal under study). But the $f_{max}$ value does not coincide with the $d_B$ of the spectrum because the interval corresponding to $\alpha(f_{max})$ does not have all the boxes that cover the spectrum. That is why $f_{max}$ is rather smaller than $d_B$.

## Hidden parameters

The most appropriate values of $\ell$ and $\Delta\alpha$ need to be found by trial and error, such that a "decent" curve $f(\alpha)$ is obtained, with $f''(\alpha) < 0$ for Winter, Summer, and Spring. For Autumn, we obtain a bi-multifractal. However, notably, for *all four seasons*, the optimal choice of $\ell$ and $\Delta\alpha$ turned out to be the same.

Another hidden parameter: The problem with the graph at each season is the meagre number of points: 106 only. Thus, we must have an artisanal way to calculate the number of points linearly interpolated between each consecutive pair (of these scant 106 points). The number of such linearly interpolating points is between 5 and 10.

### $\alpha_{max}$

Let us consider a box with very few points inside: its measure $p$ will be near zero, its $\alpha = \ln(p) / \ln(1/50)$ will be "near infinity", since $\ln(p)$ is negative and huge; $\ln 1/50$ a change of sign as before.

But ultra-light boxes do not belong to any fractal, as they might well be isolated points far away from it. They are called dispersive, and, if we take them into account, the corresponding spectrum is considerably deformed. Therefore, a maximum value of $\alpha$ has to be deduced. In order to avoid said deformities we have to truncate the α range at $\alpha = 1.40 := \alpha_{max}$.

These results are shown in Fig 10.

Observations:
1. In a practical case, e.g. the multifractal analysis of gels (Francois, Piacquadio, and Daraio, 2011), we cannot take $\ell$ "tending to zero", in fact, we cannot take it, say, smaller than the distance between atoms. The same happens to $\Delta\alpha$: a small $\Delta\alpha$, near zero, could result in a multifractal spectrum $f(\alpha) \equiv 0$!
2. The minimum number of interpolated points between adjacent rescaled Z's in a certain graph insures that the corresponding $(\alpha, f(\alpha))$ curve resembles an acceptable spectrum

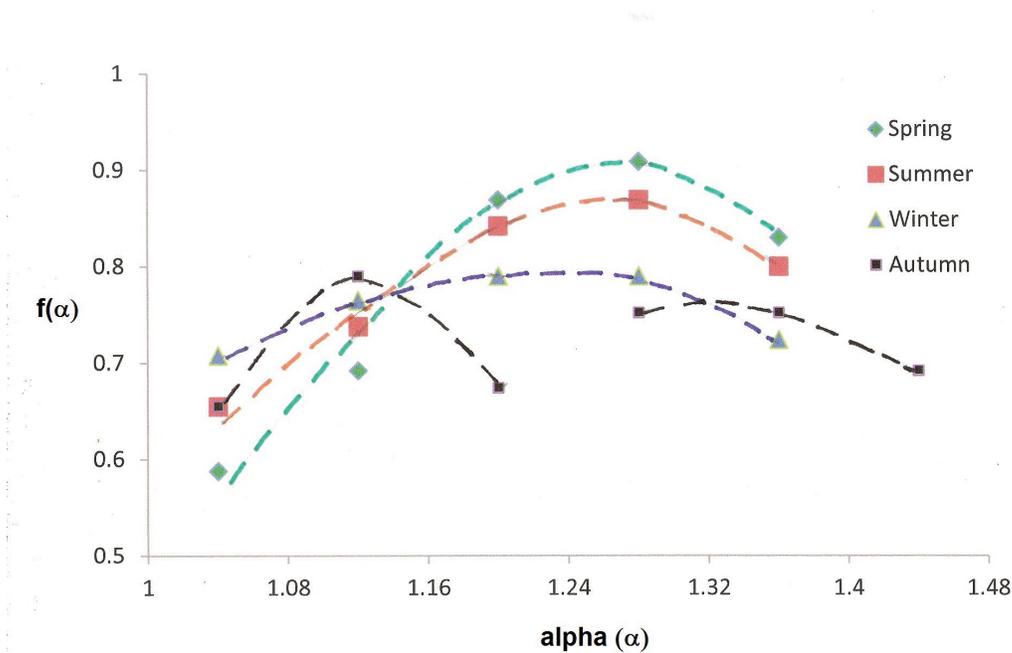

**Figure 10. The multifractal spectrum (α, f(α)) of the rescaled Z graphs corresponding to different seasons**

The spectra (α, f(α)) were determined from data collected over a century, during which the river underwent variations in precipitation, climate change, reservoirs and dams, etc. Some of these changes are reflected in the variation in rescaled Z values over the period (Fig. 9). Nevertheless, for each of three seasons we obtain a single (α, f(α)) spectrum. The spectrum corresponding to Autumn is a **bi**multifractal; as noted earlier, detailed analysis of this season should be the subject of future research.

Recall that $\alpha_{min} = 1$ and $\alpha_{max} = 1.40$. The boxes are square, $\ell = 1/50$ and $\Delta\alpha = 0.08$. Table 1 shows, for each season (except Autumn), the number of boxes in each interval of $\alpha$, the box dimension $d_B$, the Hurst coefficient, determined by $H = 2 - d_B$, as well as the maximal value of the spectral function, $f_{max}$.

Table 1. Number of boxes for each α interval, $\ell = 1/50$ and $\Delta\alpha = 0.08$, the box dimension, $d_B$, $H$, and $f_{max}$

| Intervals of α → | 1:1.08 | 1.08:1.16 | 1.16:1.24 | 1.24:1.32 | 1.32:1.40 | **Total** | $d_B$ | H | $f_{max}$ |
|---|---|---|---|---|---|---|---|---|---|
| Winter | 16 | 20 | 22 | 22 | 17 | **97** | 1.17 | 0.83 | 0.79 |
| Summer | 13 | 18 | 27 | 30 | 23 | **111** | 1.20 | 0.80 | 0.87 |
| Spring | 10 | 15 | 30 | 35 | 26 | **116** | 1.215 | 0.79 | 0.91 |

Notice that the three seasonal values of $H$, determined by $H = 2 - d_B$, are between 0.79 and 0.83, close to $H$ for the entire data set, $H_T = 0.76$, determined using the R/S method. This agrees with our initial assumption, that the seasonal values of $H$ are (roughly) the same as for the whole data series.

In Fig. 10, we see all the $f_{max}$ in the same $\Delta\alpha$. The $f_{max}$ is, as we know, smaller than the real $d_B$. The difference is a consequence of the limited size of real data sets.

**Indices for measuring variations in stream flow**

Variations in stream flow are important in the study of many phenomena: evaluation of floods and droughts, reservoir design for water supply, flood control, and hydroelectric power generation. To this list one may add climate variations—such as those associated with the El Niño Southern Oscillation (ENSO)—and climate change: the most immediate effects of climate change are changes in rainfall patterns and soil moisture content, which would affect stream flows and their seasonal variations. In order to study such variations using multifractal analysis, we propose a number of indices, described below.

Let us consider the multifractal graphs $f(\alpha)$.

A heavy box, many points $(n, Z(n))$ in it, signifies some measure of stability of the stream flow. We observe that, in the three seasons under study, the heaviest boxes, at the left of each multifractal spectrum $f(\alpha)$ are at $f_{min}$. On the other hand, light boxes, with a small number of points, at the right of the spectrum $f(\alpha)$, imply instability.

Therefore, observing the graphs, an index (say, $I_1$) of instability would be:
$I_1 = f_{max} - f_{min}$ for each season:

$$I_1^{spring} = f_{max}^{spring} - f_{min}^{spring} = 0.91 - 0.59 = 0.32$$

$$I_1^{summer} = f_{max}^{summer} - f_{min}^{summer} = 0.87 - 0.66 = 0.21$$

$$I_1^{winter} = f_{max}^{winter} - f_{min}^{winter} = 0.79 - 0.71 = 0.08$$

We observe that the most unstable situation of the Parana river stream flow at Corrientes occurs in Spring, for $I_1^{spring} > I_1^{summer} > I_1^{winter}$, Winter being the most stable season, when the flow changes little.

We want to measure this Spring instability through some other indices.

In general, the larger $I_1$ signifies <u>more instability</u> of the flow.

Since every possible Summer index that we found—in our case— (see Table 1 and values of $I_1$) seems to be intermediate between those of Spring and Winter, we will consider mainly these last two seasons.

Since heavy boxes signify stability, it makes sense to consider the number of heaviest boxes as another index:
$$I_2 := \text{\#heaviest boxes}$$

We have $I_2^{spring} = 10; < I_2^{winter} = 16; I_2^{summer} = 13$ is in between. It is obvious, again, from $I_2$ that the stream flow is least stable in Spring, and more stable in Winter: the larger $I_2$, the larger the stability.

Another index of instability is given by studying the right side of the multifractal spectrum, from $f_{max}$ to $f(\alpha_{max})$, i.e. the very light boxes. Notice that, if there are many such light boxes, their $f(\alpha)$ is large, and the difference between their $f(\alpha)$ and $f_{max}$ is small, relatively to $f_{max} - f_{min}$. Therefore, a smaller value of $I_3 := (f_{max} - f(\alpha_{max}))/I_1$ implies more instability.

$$I_3^{spring} = (0.91 - 0.83) / I_1^{spring} = 0.08 / 0.32 = 0.25$$

$$I_3^{summer} = (0.87 - 0.80) / I_1^{summer} = 0.07 / 0.21 = 0.33$$

$$I_3^{winter} = (0.79 - 0.72) / I_1^{winter} = 0.07 / 0.08 = 0.88$$

We observe that the situation in Spring is, as we know, more instable. Winter is *quite* stable.

Yet another index $I_4$ can be deduced by taking into consideration the $\alpha$-values of the spectra, rather than the $f(\alpha)$ ones.

For instance, $I_4 := (\alpha_{max} - \alpha(f_{max}))/(\alpha(f_{max}) - \alpha_{min})$, is an index, small in Spring and Summer:

$$I_4^{spring} = I_4^{summer} = (1.36 - 1.28) / (1.28 - 1.04) = 0.08/0.24 = 0.33$$

and large in Winter:
$$I_4^{winter} = \frac{1.36 - 1.24}{1.24 - 1.04} = \frac{0.12}{0.20} = 0.60$$

The smaller $I_4$ is, the larger the instability.

The values of the indices for each season are summarized in Table 2.

**Table 2. Indices, for Spring, Summer and Winter**

| Season | $I_1$ | $I_2$ | $I_3$ | $I_4$ |
|---|---|---|---|---|
| Spring | 0.32 | 10 | 0.25 | 0.33 |
| Summer | 0.21 | 13 | 0.33 | 0.33 |
| Winter | 0.08 | 16 | 0.88 | 0.60 |

All the indices $I_1$, $I_2$, $I_3$, and $I_4$ indicate that Winter is the most stable season; Summer is next according to three of these indices ($I_1$, $I_2$, and $I_3$) followed by Spring; for $I_4$, Spring and Summer, are about the same. As we noted earlier, the Boxplot of seasonal stream flow data shows roughly the same inter-quartile range for each season, and indeed, Winter shows the largest variation. Thus, these indices provide insight into river flow stability, not detected in—and indeed contradictory to—that from basic statistical analysis.

The indices may be compared with the Rescaled Z for each season. The vertical ranges of rescaled Z values are directly linked to the indices that reflect the stability of each season.

Leaving aside Autumn, which has a bimultifractal structure, the range of rescaled Z for Winter is the smallest of the three remaining seasons; this coincides with the conclusions of the indices: Winter is the most stable and Spring the least. While the values of the indices allow a quantification of the stability of each season, the latter can be immediately seen from the shapes of the $(\alpha, f(\alpha))$ spectra in Figure 10.

## Conclusions

The Hurst coefficient of stream flow over the 106-year period was found to be 0.76, a value within the wide range obtained for different rivers in other studies.

We select the data corresponding to the flow of four different seasons. The rescaled Z showed strong differences across seasons: the range of rescaled Z is small for Autumn, with no long-term trends. For the other three seasons, Winter has the smallest range, followed by Summer and finally by Spring (Figure 9).

For three of the four seasons, the rescaled Z values decline up to a certain point (1978 for Winter and 1972 for Spring and Summer) and increase thereafter. Thus, the rescaled Z values reveal trends not visible in the daily stream flow data (Figure 2).

Multifractal spectra for each season were determined by the box-counting method. The optimal choice of the box size, defined by $\ell$ and $\Delta\alpha$, was found to be the same for all four seasons. For three of the four seasons (Winter, Summer, and Spring), we obtain smooth curves with $f''(\alpha) < 0$, despite variations in hydrology over the 106 years. For Autumn, we obtain a bi-multifractal. The latter form would not emerge for a multifractal spectrum determined using the thermodynamic algorithm, which always forces $f''(\alpha)$ to be $< 0$. This confirms our experience on the limitations of the thermodynamic algorithm.

FBM and multifractal analysis yielded conclusions different from, and sometimes contradictory with, the results of basic statistical analysis, especially with respect to seasonal stability.

A number of indices, derived from the geometry of the multifractal spectra, allow a comparison of the stream flow behavior in the different seasons.

## Acknowledgements

We thank Ricardo Arias, Department of Electronics, School of Engineering, University of Buenos Aires, for his participation in early parts of this study.